\documentclass[aps,prl,preprint,showpacs]{revtex4}
\usepackage{amsfonts}
\usepackage{amsmath}
\usepackage{mathrsfs}
\begin{document}
\title{Physical decomposition of the gauge and gravitational fields}

\author{Xiang-Song Chen$^{1,2,3,}$}
\email{cxs@hust.edu.cn}
\author{Ben-Chao Zhu$^1$}

\affiliation{$^1$Department of Physics, Huazhong
University of Science and Technology, Wuhan 430074, China\\
$^2$ Joint Center for Particle, Nuclear Physics and Cosmology,
Nanjing 210093, China \\
$^3$Kavli Institute for Theoretical Physics China, CAS, Beijing
100190, China}

\date{\today}

\begin{abstract}
Physical decomposition of the non-Abelian gauge field has recently
solved the two-decade-lasting problem of a meaningful gluon spin.
Here we extend this approach to gravity and attack the
century-lasting problem of a meaningful gravitational energy. The
metric is unambiguously separated into a pure geometric term which
contributes null curvature tensor, and a physical term which
represents the true gravitational effect and always vanishes in a
flat space-time. By this decomposition the conventional
pseudo-tensors of the gravitational stress-energy are easily rescued
to produce definite physical result. Our decomposition applies to
any symmetric tensor, and has interesting relation to the
transverse-traceless (TT) decomposition discussed by Arnowitt, Deser
and Misner, and by York.

\pacs{11.15.-q, 04.20.Cv}
\end{abstract}
\maketitle

Gauge invariance is the most elegant and efficient principle for
constructing interactions in the present field theories of physics.
By requiring field equations to be gauge invariant, the manner of
the couplings (and self-couplings) of various fields are strongly
constrained. This applies both to the standard model of the strong
and electro-weak interactions, and to Einstein's gravitational
theory. For the latter case gauge invariance refers to general
covariance under arbitrary coordinate transformation. It is rather
annoying, however, that a theory built uniquely out of the
gauge-invariance requirement does not seem to guarantee gauge
invariance for all physical quantities. In hadron physics, e.g., in
the two-decade efforts to understand how the nucleon spin originates
from the spin and orbital motion of its quark and gluon
constituents, one encounters severe difficulty in finding a
gauge-invariant description of the gluon spin and orbital angular
momentum. Only recently, a solution was obtained in Ref.
\cite{Chen08}, and further developed in Ref. \cite{Chen09}. A more
celebrated and still unsolved gauge-dependence problem is the energy
density of the gravitational field. After countless attempts of
nearly a century, a convincing solution is still lacking. A
reflection of this desperation is the often heard argument that,
since the effect of gravity at any point can be eliminated by
transiting to a free-fall frame, gravitational energy is
intrinsically non-localizable and can at best be quasi-local to a
closed two-surface \cite{Yau09,Szab09}.

The key obstacle to constructing all physical quantities
gauge-invariantly is the inevitable involvement of the gauge or
gravitational field together with their ordinary derivatives, which
are all intrinsically gauge dependent. The idea in Refs.
\cite{Chen08,Chen09} is to decompose the gauge field: $A_\mu\equiv
\hat A_\mu+\bar A_\mu$. The aim is that $\hat A_\mu$ will be a
physical term which is gauge-covariant and always vanishes in the
vacuum, and $\bar A_\mu$ is a pure-gauge term which solely carries
the gauge freedom and has no essential physical effects
(particularly, it does not contribute to the electric or magnetic
field strength). Equipped with the separate $\hat A_\mu$ and $\bar
A_\mu$, a naively gauge-dependent quantity (such as the gluon spin
$\vec S =\vec E\times \vec A$) can easily be rescued to be
gauge-covariant, simply by replacing $A_\mu$ with $\hat A_\mu$, and
by replacing the ordinary derivative with the pure-gauge covariant
derivative constructed with $\bar A_\mu$ instead of $A_\mu$.

Mathematically, a well-defined separation $A_\mu =\hat A_\mu + \bar
A_\mu$ means an unambiguous prescription for constructing $\hat
A_\mu$ and $\bar A_\mu$ out of a given $A_\mu$. The properties
(especially, gauge transformations) of $\hat A_\mu$ and $\bar A_\mu$
are then inherently determined via their mathematical expressions in
terms of $A_\mu$. In Refs. \cite{Chen08,Chen09}, it was found that
$\hat A_\mu$ and $\bar A_\mu$ can indeed be solved in terms of
$A_\mu$ by setting up proper differential equations and boundary
conditions, which lead to unique solutions for $\hat A_\mu$ and
$\bar A_\mu$ with desired physical properties. In this paper, we
show that this method can be generalized to gravitational theory.
The metric tensor $g_{\mu\nu}$ is unambiguously decomposed into the
sum of a physical term $\hat g_{\mu\nu}$, which represents the true
gravitational effect, and a pure geometric term $\bar g_{\mu\nu}$,
which represents the spurious gravitational effect associated with
coordinate choice. Gauge-dependence of the gravitational energy
originates exactly from the fact that the metric may contain a
spurious gravitational effect. While in a flat space-time the
Cartesian coordinate with vanishing affine connection seems a
natural choice, in an intrinsically curved space-time no coordinate
is obviously more natural than others, hence it is no longer a
trivial task to get rid of the spurious gravitational effect. In an
accompanying paper \cite{Chen10}, we discuss a gauge-fixing
approach, by defining a unique physical coordinate which contains no
spurious gravitational effect. In this paper, we present the more
general field-decomposition approach, by seeking a prescription to
identify the geometric $\bar g_{\mu\nu}$ for a given metric $
g_{\mu\nu}$ in any coordinate.

As for gauge theories, we find that the prescription is again a set
of defining differential equations, which are displayed most
concisely in the form \cite{note1}:
\begin{subequations}
\label{G}
\begin{eqnarray}
\bar R^\rho_{~\sigma \mu\nu} &\equiv &\partial_\mu
\bar\Gamma^\rho_{\sigma \nu}-\partial _\nu \bar\Gamma ^\rho_{\sigma
\mu} +\bar\Gamma ^\rho_{\alpha \mu}\bar\Gamma ^\alpha _{\sigma \nu}
-\bar\Gamma ^\rho_{\alpha \nu} \bar\Gamma ^\alpha _{\sigma \mu} =0,
\label{G1}
\\
g^{ij}\hat \Gamma ^\rho_{ij}&=&0 . \label{G2}
\end{eqnarray}
\end{subequations}
The notations require some caution: $\bar \Gamma ^\rho_{\mu\nu}$ is
the purely geometric part of the affine connection. Its relation to
$\bar g_{\mu\nu}$ is analogous to that of $\Gamma ^\rho_{\mu\nu}$
and $g_{\mu\nu}$:
\begin{equation}
\bar \Gamma ^\rho_{\mu\nu}\equiv \frac 12 \bar g^{\rho\sigma}
(\partial_\mu \bar g_{\sigma \nu} +\partial_\nu \bar g_{\sigma\mu}
-\partial_\sigma \bar g_{\mu\nu}). \label{barg}
\end{equation}
Here $\bar g^{\mu\nu}$ is defined as the inverse of $\bar
g_{\mu\nu}$, i.e., $\bar g^{\mu\rho} \bar g_{\rho\nu } =\delta
^\mu_{~\nu}$. The aim of this choice is that $\bar R^\rho_{~\sigma
\mu\nu}$ in (\ref{G1}) is just the Riemann curvature of $\bar
g_{\mu\nu}$. It must then be noted that $\hat g^{\mu\nu}\equiv
g^{\mu\nu} -\bar g^{\mu\nu}$ is not the inverse of $\hat
g_{\mu\nu}$. (In fact, the physical term $\hat g_{\mu\nu}$ may not
have an inverse at all.) The difference $\Gamma ^\rho_{\mu\nu}-\bar
\Gamma ^\rho_{\mu\nu} \equiv \hat \Gamma ^\rho_{\mu\nu}$ is defined
as the physical connection. It is not related to $\hat g_{\mu\nu}$
as in Eq. (\ref{barg}).

To comprehend how Eq. (\ref{G}) is chosen, how it gives solution for
$\hat \Gamma^\rho_{\sigma\mu}$ and $\bar \Gamma^\rho_{\sigma\mu}$
[and further for $\hat g_{\mu\nu} $ and $\bar g_{\mu\nu}$ by Eq.
(\ref{barg})] with desired properties, and how the solution in turn
is employed to solve the gauge-dependence problem of the
gravitational energy, it is most helpful to recall the parallel
constructions for gauge theories in Refs. \cite{Chen08,Chen09}. In
Abelian case, the gauge field $A_\mu$ transforms as $A_\mu
\rightarrow A'_\mu = A_\mu -
\partial_\mu \omega$, which leaves the field strength invariant:
$F_{\mu\nu}=\partial _\mu A_\nu-\partial_\nu A_\mu \rightarrow
F'_{\mu\nu}=F_{\mu\nu}$. The defining equations for the separation
$A_\mu =\hat A_\mu + \bar A_\mu$ are
\begin{subequations}
\label{A}
\begin{eqnarray}
\bar F_{\mu\nu} &\equiv &\partial _\mu \bar A_\nu-\partial_\nu \bar
A_\mu=0 ,\label{A1}\\
\partial_i \hat A_i &=&0 . \label{A2}
\end{eqnarray}
\end{subequations}

Eq.~(\ref{A1}) has very clear physical meaning: the pure-gauge term
$\bar A_\mu$ gives null field strength. Eq.~(\ref{A2}) can be
regarded as the transverse condition for a physical photon with zero
mass. But to avoid confusion with the radiation gauge condition
$\partial _i A_i=0$ for the full $A_i$, it is more helpful to think
in a mathematical way that Eq. (\ref{A}) are the needed differential
equations to solve $\hat A_\mu$ and $\bar A_\mu$. Since $\hat
A_\mu+\bar A_\mu=A_\mu$, it suffices to examine $\hat A_\mu$. To
this end we rewrite Eq. (\ref{A1}) as
\begin{equation}
\partial _\mu \hat A_\nu-\partial_\nu \hat
A_\mu=F_{\mu\nu}. \tag{\ref{A1}$'$} \label{A1'}
\end{equation}
A clever way to solve is to act on both sides with $\partial_i$, set
$\mu=i$ and sum over $i$, and use Eq. (\ref{A2}). This gives
\begin{equation}
\vec \nabla ^2 \hat A_\nu =\partial _i F_{i\nu}, ~{\rm or}~ \hat
A_\nu = \frac 1{\vec\nabla ^2} \partial_i F_{i\nu}, \label{As}
\end{equation}
where we have required a natural boundary condition that, for a
finite system, the physical term $\hat A_\mu$ vanish at infinity, as
does the field strength $F_{\mu\nu}$. \cite{noteBound} The explicit
solution in Eq. (\ref{As}) indicates clearly that the physical field
$\hat A_\mu$ is gauge invariant, and hence the pure-gauge field
$\bar A_\mu =A_\mu -\hat A_\mu$ carries all the gauge freedom and
transforms in the same manner as does the full $A_\mu$. Moreover,
Eq.~(\ref{As}) tells us that the physical term $\hat A_\mu $
vanishes if the field strength $F_{\mu\nu} = 0$.

In non-Abelian case, the gauge transformation is more complicated:
$A'_\mu = U A_\mu U^\dagger -\frac ig U\partial_\mu U^\dagger$. The
field strength now contains a self-interaction term, and transforms
covariantly instead of invariantly: $F_{\mu\nu}=
\partial _\mu A_\nu-\partial_\nu A_\mu+ig[A_\mu,A_\nu]
\rightarrow F'_{\mu\nu}= U F_{\mu\nu} U^\dagger$. It is fairly
non-trivial to choose proper defining equations for the non-Abelian
$\hat A_\mu$ and $\bar A_\mu$. They were originally proposed in Ref.
\cite{Chen08}, and further developed in Ref. \cite{Chen09} to be:
\begin{subequations}
\label{nA}
\begin{eqnarray}
\bar F_{\mu\nu} &\equiv &\partial _\mu \bar A_\nu-\partial_\nu \bar
A_\mu +ig[\bar A_\mu,\bar A_\nu]=0 , \label{nA1}\\
\bar {\mathcal D}_i \hat A_i &\equiv& \partial_i \hat A_i +ig[\bar
A_i,\hat A_i]= 0 .\label{nA2}
\end{eqnarray}
\end{subequations}
We will shortly show that Eq. (\ref{nA}) gives solution for $\hat
A_\mu$ and $\bar A_\mu$ with desired gauge-transformation
properties:
\begin{equation}
\hat A'_\mu =  U \hat A_\mu U^\dagger, ~~~ \bar A'_\mu =U \bar A_\mu
U^\dagger -\frac ig U\partial_\mu U^\dagger . \label{T}
\end{equation}
By these properties, $\bar{\mathcal D}_\mu=\partial_\mu +ig[\bar
A_\mu,~]$ is a pure-gauge covariant derivative for the adjoint
representation, and Eq. (\ref{nA}) is covariant under non-Abelian
gauge transformations. Analogous to the Abelian case, Eq. (\ref{nA})
says that $\bar A_\mu$ is a pure-gauge field giving null field
strength, and the physical field $\hat A_\mu$ satisfies a
``covariant transverse condition''. However, as we remarked in the
Abelian case, the real justification for Eq.~(\ref{nA}) is that they
are the right mathematic equations to solve $\hat A_\mu$ and $\bar
A_\mu$ in terms of $A_\mu$, with desired gauge transformations in
(\ref{T}). Again, we examine $\hat A_\mu$ with trivial boundary
condition, and rewrite Eq.~(\ref{nA}):
\begin{subequations}
\label{nA'}
\begin{eqnarray}
\partial _\mu \hat A_\nu-\partial_\nu \hat
A_\mu & = &F_{\mu\nu} + ig([\hat A_\mu -A_\mu,\hat A_\nu]-[\hat
A_\mu,
A_\nu]) \label{nA1'} \\
\partial_i \hat A_i &= &ig[\hat A_i, A_i]. \label{nA2'}
\end{eqnarray}
\end{subequations}
Due to non-linearity, these are not easy to solve. To proceed, we
employ the usual technique of perturbative expansion, which applies
when either the coupling constant $g$ or the field amplitude is
small. For a small $g$, e.g., we write $\hat A_\mu =\hat A_\mu^{(0)}
+g \hat A_\mu^{(1)} + g^2 \hat A_\mu^{(2)} + \cdots$. Eq.
(\ref{nA'}) can then be solved order by order. The zeroth-order term
$\hat A_\mu ^{(0)}$ satisfy the same equations as (\ref{A1'}) and
(\ref{A2}). Its solution is given by Eq.~(\ref{As}), and can in turn
be used to solve the equations for the leading non-trivial term
$\hat A_\mu^{(1)}$:
\begin{subequations}
\label{nAg}
\begin{eqnarray}
\partial _\mu \hat A_\nu^{(1)}-\partial_\nu \hat A_\mu^{(1)} &=&
i[\hat A_\mu^{(0)}-A_\mu, \hat A_\nu^{(0)}] -i[\hat A_\mu^{(0)}, A_\nu] ,\\
\partial_i \hat A_i^{(1)}&=& i[\hat A_i^{(0)}, A_i] .
\end{eqnarray}
\end{subequations}
The solution is obtained by the same strategy for Abelian case, and
can be further employed to solve the next-order term $\hat
A_\mu^{(2)}$, and so on. Given validity of this perturbative
expansion, the solution to Eq.~(\ref{nA'}) is unique. This
uniqueness has important implications: a) $F_{\mu\nu} =0$ is
necessary and sufficient for $\hat A_\mu =0$; and b) $\hat A_\mu$
and $\bar A_\mu$ have the gauge transformations as in (\ref{T}). The
proof of b) is as follows: Eq. (\ref{T}) is solution of Eq.
(\ref{nA}) with $\hat A_\mu$ and $\bar A_\mu$ replaced by $\hat
A'_\mu$ and $\bar A'_\mu$, and since the solution to Eq. (\ref{nA})
is unique, Eq. (\ref{T}) gives the right gauge transformations.

We now turn to the gravitational equations (\ref{G}) and
({\ref{barg}). Because of non-linearity, we have to rely again on
perturbative method, and require that the gravitational field be at
most moderately strong. Namely, the magnitude of $h_{\mu\nu}\equiv
g_{\mu\nu}- \eta_{\mu\nu}$ (with $\eta_{\mu\nu}$ the Minkowski
metric) is smaller than 1 and can be treated as an expansion
parameter. It then takes a little algebra to show that Eqs.
(\ref{G}) and ({\ref{barg}) can be solved similar to the gauge-field
equations. We proceed by first looking at the physical connection
$\hat \Gamma ^\rho_{\sigma\nu}$, to which we can assign a natural
boundary condition that (for a finite system) $\hat \Gamma ^\rho
_{\sigma \nu}$ vanish at infinity as does the Riemann curvature $R
^\rho _{~\sigma \mu\nu}$. We define an expansion $\hat \Gamma ^\rho
_{\sigma \nu}=\hat \Gamma ^{\rho(1)} _{\sigma \nu} +\hat \Gamma
^{\rho(2)} _{\sigma \nu} +\cdots $ in orders of $h_{\mu\nu}$. For
the first-order term $\hat \Gamma ^\rho _{\sigma \nu}$, we get from
Eq. (\ref{G})
\begin{subequations}
\label{Gh}
\begin{eqnarray}
\partial_\mu \hat\Gamma^{\rho(1)}_{\sigma \nu}-\partial _\nu \hat\Gamma
^{\rho(1)}_{\sigma \mu} & = & R^\rho_{~\sigma \mu\nu}, \label{Gh1}
\\
\hat \Gamma ^{\rho(1)} _{ii} & = & 0. \label{Gh2}
\end{eqnarray}
\end{subequations}
{\it Solution:} Set $\mu=\sigma =i$ in Eq. (\ref{Gh1}), sum over
$i$, and use Eq. (\ref{Gh2}), we get
\begin{equation}
\partial_i \hat\Gamma^{\rho(1)}_{i \nu}=R^\rho_{~ii\nu}. \label{Gh2'}
\end{equation}
Then, act on both sides of Eq. (\ref{Gh1}) with $\partial_i$, set
$\mu=i$, sum over $i$, and use Eq. (\ref{Gh2'}), we obtain the
solution
\begin{equation}
\hat \Gamma ^{\rho(1)} _{\sigma \nu} = \frac 1{\vec\nabla ^2}
(\partial_i R ^\rho _{~\sigma i\nu} + \partial_\nu R^\rho
_{~ii\sigma}). \label{Gs1}
\end{equation}
This can then be employed to solve the second-order term $\hat
\Gamma ^{\rho(2)}_{\sigma \nu}$. From Eq. (\ref{G}), we have
\begin{subequations}
\label{Ghh}
\begin{eqnarray}
\partial_\mu \hat\Gamma^{\rho(2)}_{\sigma \nu} -\partial _\nu
\hat\Gamma ^{\rho(2)}_{\sigma \mu} &=&(\hat\Gamma ^{\rho(1)}_{\alpha
\mu} -\Gamma ^\rho_{\alpha \mu})\hat\Gamma ^{\alpha(1)} _{\sigma
\nu} -\hat \Gamma ^{\rho(1)}_{\alpha \mu}\Gamma ^\alpha _{\sigma
\nu} -(\hat\Gamma ^{\rho(1)}_{\alpha \nu} -\Gamma ^\rho_{\alpha
\nu})\hat\Gamma ^{\alpha (1)}_{\sigma \mu} +\hat\Gamma
^{\rho(1)}_{\alpha \nu}\Gamma ^\alpha _{\sigma \mu},
\\
\hat \Gamma ^{\rho(2)} _{ii} &=&h^{ij} \hat \Gamma ^{\rho(1)} _{ij}.
\end{eqnarray}
\end{subequations}
Here $h^{\mu\nu}\equiv \eta^{\mu\nu} -g^{\mu\nu}$. Though looking
tedious, Eq. (\ref{Ghh}) can be solved similar to Eq. (\ref{Gh}).
The solution can be further employed to continue the perturbative
procedure up to any desired order, in principle.

Having separated the affine connection, we can use Eq. (\ref{barg})
to solve the metric separation, $g_{\mu\nu}\equiv \bar
g_{\mu\nu}+\hat g_{\mu\nu}$. It is useful to define
$h_{\mu\nu}\equiv \bar h_{\mu\nu} +\hat h_{\mu\nu}$, thus $\bar
g_{\mu\nu}=\eta_{\mu\nu}+\bar h_{\mu\nu}$ and $\hat g_{\mu\nu}=\hat
h_{\mu\nu}$. We again look at the physical term $\hat h_{\mu\nu}$
which can be assigned a trivial boundary condition. As for $\hat
\Gamma ^\rho_{\mu\nu}$, we define an expansion $\hat
h_{\mu\nu}\equiv \hat h_{\mu\nu}^{(1)} +\hat
h_{\mu\nu}^{(2)}+\cdots$ in orders of $h_{\mu\nu}$. From Eq.
(\ref{barg}), we derive the first-order equation
\begin{equation}
\partial _\mu
\hat h_{\sigma \nu}^{(1)} +\partial_\nu \hat h_{\sigma
\mu}^{(1)}-\partial_\sigma \hat h _{\mu\nu} ^{(1)}=2\eta
_{\rho\sigma}\hat \Gamma ^{\rho(1)} _{\mu\nu} \label{hatg}
\end{equation}
Interchange $\sigma, \nu$ in Eq. (\ref{hatg}) and add the result
back to Eq. (\ref{hatg}), we get
\begin{equation}
\partial _\mu \hat h_{\sigma \nu}^{(1)} =\eta
_{\rho\sigma} \hat  \Gamma ^{\rho(1)} _{\mu\nu}+\eta _{\rho\nu} \hat
\Gamma ^{\rho(1)} _{\mu\sigma}.   \label{hatg'}
\end{equation}
Act on both sides with $\partial_i$, set $\mu=i$ and sum over $i$,
we obtain
\begin{equation}
\vec\nabla ^2 \hat h_{\sigma \nu} ^{(1)}=\partial_i (\eta
_{\rho\sigma} \hat \Gamma  ^{\rho(1)} _{i\nu}+\eta _{\rho\nu} \hat
\Gamma  ^{\rho(1)} _{ i\sigma}) = \eta _{\rho\sigma} R ^\rho
_{~ii\nu}+\eta _{\rho\nu} R^\rho _{~i i\sigma}.
\end{equation}
where in the second step we have used Eq. (\ref{Gh2'}). Since this
is the first-order equation, indices can be lowered by the Minkowski
metric. Then by noticing the symmetry property of $R_{\sigma
ii\nu}$, we finally obtain the solution
\begin{eqnarray}
\hat h_{\sigma \nu} ^{(1)}&=&2\frac 1{\vec\nabla ^2} R_{\sigma
ii\nu}^{(1)}
\nonumber \\
&=&h_{\sigma \nu} - \frac 1{\vec\nabla ^2}(h_{\nu
i,i\sigma}+h_{\sigma i,i\nu} -h_{ii,\sigma\nu}). \label{hatgs}
\end{eqnarray}
Here and below a comma is used to denote derivative when too many
occur. The superscript on $R_{\sigma ii\nu}^{(1)}$ is to remind that
it is computed to first-order in $h_{\mu\nu}$. Rigorously speaking,
the second expression requires that $h_{\mu\nu}$ (not just
$R^\rho_{~\sigma\mu\nu}$) vanish at infinity.

For the second-order term $\hat h_{\sigma \nu} ^{(2)}$, we derive
from Eq. (\ref{barg})
\begin{eqnarray}
\hat g_{\sigma \nu,\mu}^{(2)} +\hat g_{\sigma \mu,\nu}^{(2)}- \hat g
_{\mu\nu,\sigma}^{(2)}=2\eta _{\rho\sigma}\hat \Gamma ^{\rho(2)}
_{\mu\nu}+ \eta^{\alpha \rho} \hat h^{(1)}_{\rho \sigma}( h_{\alpha
\nu,\mu}+ h_{\alpha \mu,\nu} -  h_{\mu \nu,\alpha}) \nonumber
\\+\eta^{\alpha \rho}(h_{\rho\sigma}- \hat h^{ (1)}_{\rho \sigma})(
\hat h_{\alpha \nu,\mu}^{(1)}+ \hat h_{\alpha \mu,\nu}^{(1)} - \hat
h_{\mu \nu,\alpha}^{(1)}). \label{hatg2}
\end{eqnarray}
Solution of $\hat h_{\sigma \mu}^{(2)}$ is similar to $\hat
h_{\sigma \mu}^{(1)}$, though more tedious. The perturbative
solution for $\hat h_{\mu\nu}$ can be continued to the same order as
$\hat \Gamma ^\rho_{\mu\nu}$.

After obtaining $\hat g_{\mu\nu}$ and $\bar
g_{\mu\nu}=g_{\mu\nu}-\hat g_{\mu\nu}$, we must remark on how $\bar
g^{\mu\nu}$ and $\hat g^{\mu\nu}$ are computed. By definition, $\bar
g^{\mu\nu}$ is the inverse of $\bar g_{\mu\nu}$. Then, $\hat
g^{\mu\nu}$ is computed as $g^{\mu\nu}-\bar g^{\mu\nu}$. At lowest
order, $h^{\mu\nu}$, $\bar g^{\mu\nu}$, and $\hat g^{\mu\nu}$ are
just related to $h_{\mu\nu}$, $\bar g_{\mu\nu}$, and $\hat
g_{\mu\nu}$ by the Minkowski metric. But this property is lost at
higher orders.

The solutions we obtain show the desired property that the physical
terms $\hat \Gamma ^\rho_{\sigma \mu}$ and $\hat h_{\mu\nu}$ vanish
if and only if $R^\rho_{~\sigma \mu\nu}=0$, i.e., the space-time is
intrinsically flat. It is also illuminating to look at the property
of the pure geometric terms $\bar \Gamma ^\rho_{\sigma \mu}$ and
$\bar h_{\mu\nu}$. To this end we rewrite Eq.~(\ref{G2}) as
$g^{ij}\bar\Gamma ^\rho_{ij}=g^{ij}\Gamma ^\rho_{ij}$. This
indicates that in order to have $\bar \Gamma ^\rho_{\sigma \mu}=0$
(so that the spurious gravitational effect is absent), it is
necessary that $g^{ij}\Gamma ^\rho_{ij}=0$. On the other hand, given
validity of our perturbative expansion, $g^{ij}\Gamma ^\rho_{ij}=0$
will lead uniquely to $\bar \Gamma ^\rho_{\sigma \mu}=0$. We
therefore name a coordinate in which $g^{ij}\Gamma ^\rho_{ij}=0$ the
``pertinent coordinate''. (Similarly, in gauge theories, the
radiation gauge $\partial_i A_i=0$ leads to the solution for the
pure-gauge field $\bar A_\mu =0$, and can be termed the ``pertinent
gauge'' \cite{note2}.) The pertinency condition $g^{ij}\Gamma
^\rho_{ij}=0$ is just what we find in Ref. \cite{Chen10} the ``true
radiation gauge for gravity''. It is straightforward to verify that
the spherical coordinate is not ``pertinent'' even in a flat
space-time. This explains why it gives unreasonable gravitational
energy by the traditional pseudo-tensors.

We note that the pertinency condition is fairly non-trivial. E.g.,
while the Cartesian coordinate in flat space-time gives $\Gamma
^\rho _{\mu\nu}\equiv 0$ and is clearly pertinent, the
quasi-Cartesian coordinate in a curved space-time is not necessary
pertinent, e.g., the simplest Schwarzschild solution:
$ds^2=\left(\frac{1-MG/2r}{1+MG/2r}\right)^2 dt^2
-\left(1+\frac{MG}{2r}\right)^4 d\vec r ^2$. Moreover, it is not
trivial to convert this coordinate to a pertinent one, except at
linear order \cite{Chen10}. It is exactly the non-triviality of the
pertinency condition that calls for our field-decomposition
approach, which works straightforwardly in any coordinate, and can
pick out the true gravitational content of the metric up to moderate
strength.

We are now in the position to explain how to calculate a physically
meaningful energy density of the gravitational field, for any given
$g_{\mu\nu}$ of a finite and not-too-strong gravitating system. The
metric $g_{\mu\nu}$ may either be obtained by solving the Einstein
equation directly, or may just be worked out with some guessing, or
even be the experimentally measured result. First, the metric is put
into the pertinency test: If one finds $g^{ij}\Gamma ^\rho_{ij}=0$,
it means that this $g_{\mu\nu}$ contains no spurious gravitational
effect, thus can be used directly in the traditional pseudo-tensors
to compute the energy density. If, instead, $g^{ij}\Gamma
^\rho_{ij}\neq 0$, it means that this $g_{\mu\nu}$ does contain
spurious gravitational effect, and one should revise a pseudo-tensor
by replacing the quantities in it with their corresponding physical
counterparts, which are obtained by the field-decomposition approach
we just presented. This would give a concrete gravitational energy
as physical as that in the pertinent coordinate.

\textit{Discussion.}---(i) Various pseudo-tensors show a high
degeneracy concerning the total energy of a gravitating body. It
would be interesting to examine whether such degeneracy persists to
the level of a meaningful density.

(ii) In gauge theories, gauge transformation and Lorentz
transformation are two different manipulations. Therefore, in
Eq.~(\ref{A2})/(\ref{nA2}), $\hat A_\mu$ is gauge
invariant/covariant so as to make the equation gauge
invariant/covariant. However, to make the equation hold in any
Lorentz frame, the physical field $\hat A_\mu$ must not transform as
a four-vector. This is an inevitable physical feature of a massless
particle with spin-1 or higher \cite{Wein95}. In general relativity,
however, gauge transformation and coordinate transformation mean the
same thing. Therefore, to make Eq.~(\ref{G2}) hold in any
coordinate, the physical term $\hat \Gamma ^\rho_{\sigma\mu}$ must
not transform covariantly under four-dimensional transformations,
even linear (Lorentz) ones. This manifests the masslessness of the
gravitational field. But by our construction $\hat \Gamma
^\rho_{\sigma\mu}$ is indeed a true tensor under spatial
transformations, following the same line as in proving the
non-Abelian transformations in Eq. (\ref{T}).

(iii) At leading order, $h^{(1)}_{\mu\nu}$ is essentially the field
defined in the ``pertinent coordinate'' as we discuss in Ref.
\cite{Chen10}, where we have derived the second expression in Eq.
(\ref{hatgs}) by a method of gauge transformation. Moreover, the
expression mimics exactly the form of the ``transverse'' part of the
matter stress-energy tensor, derive in Ref. \cite{Chen10} by yet
another method:
\begin{equation}
\hat S_{ij}=S_{ij}-\frac {1}{\vec \nabla ^2} (\partial_i\partial_k
S^k_{~j}+\partial_j \partial_k S^k_{~i} -\partial_i\partial_j
S^k_{~k}), \label{hatS}
\end{equation}
where $S_{\mu\nu} \equiv T_{\mu\nu}-\frac 12 \eta_{\mu\nu}
T^\rho_{~\rho}$. This ``coincidence'' is actually profound and
reveals that our tensor-separation is a unique extension of the
usual vector-separation by curl-free and divergence-free conditions:
Riemann curvature is the unique covariant ``curl'' of a tensor,
hence comes Eq. (\ref{G1}). The uniqueness of the expression in Eq.
(\ref{G2}) is explained in \cite{Chen10}.

(iv) Arnowitt, Deser and Misner (ADM) discussed a linear orthogonal
separation of a symmetric spatial tensor \cite{ADM}: $h_{ij} =
h_{ij}^{TT}+h_{ij}^T+h_{ij}^{L}$, where $h_{ij}^{TT}$ is transverse
and traceless, $h_{ij}^T$ is transverse, and $h_{ij}^{L}$ is
longitudinal; all expressed uniquely via $h_{ij}$:
\begin{subequations}
\label{TTD}
\begin{eqnarray}
h_{ij}^L&=&f_{i,j}+ f_{j,i},~f_i= \frac 1{\vec \nabla ^2}
(h_{ik,k} -\frac 12 \frac 1{\vec \nabla ^2} h_{kl,kli}) \\
h_{ij}^T &=&\frac 12 (\delta _{ij} f^T -\frac 1{\vec \nabla ^2}
f^T_{,ij}), ~ f^T =h_{kk}-\frac 1{\vec \nabla ^2} h_{kl,kl}\\
h_{ij}^{TT}&=&h_{ij} - h_{ij}^T-h_{ij}^{L} .
\end{eqnarray}
\end{subequations}
ADM regard $h_{ij}^{TT}$ as the physical part of the gravitational
field. At linear order, both $h_{ij}^{TT}$ and our $\hat h_{\mu\nu}$
are gauge invariant. But a key difference is that in our method the
rest part $\bar h_{\mu\nu}$ is a pure gauge, while in the ADM method
$h^T_{ij}$ is also gauge invariant and only $h^L_{ij}$ is a pure
gauge. This implies that $h_{ij}^{TT}$ does not contain all physical
content of $h_{ij}$, and is not as pertinent as $\hat h_{\mu\nu}$.
Since at linear order $\hat h_{\mu\nu}$, $h_{ij}^{TT}$, and
$h^T_{ij}$ are all gauge-invariant, we can expect some relations
among them. Remarkably, indeed, a little algebra shows
\begin{subequations}
\begin{eqnarray}
f^T &=&\frac 12 \hat h ^{(1)}_{kk}=h_{kk}-\frac 1{\vec \nabla ^2}
h_{kl,kl}, \\
h_{ij}^{TT}&=&\hat h^{(1)}_{ij} - \frac 14 (\delta _{ij} \hat h
^{(1)}_{kk} +\frac 1{\vec \nabla ^2} \hat h ^{(1)}_{kk,ij}).
\label{h-TT}
\end{eqnarray}
\end{subequations}
Thus, the relation of $h_{ij}^{TT}$ and $\hat h_{ij}$ is similar to
that of the TT gauge and our pertinency condition: They agree for
pure waves without matter source, but disagree otherwise
\cite{Chen10}.

(v) York has proposed a different extraction of TT component from a
symmetric tensor: $h^{TT}_{ij}\equiv h_{ij}-\tilde h^L_{ij}-\frac 13
\delta _{ij} h_{kk}$, with $\tilde h^L_{ij}$ another longitudinal
part and $\frac 13 \delta _{ij} h_{kk}$ a trace part. \cite{York} At
linear order, the explicit expression is:
\begin{subequations}
\begin{eqnarray}
\tilde h^L_{ij} &=&W_{i,j} +W_{j,i} -\frac 23 \delta _{ij} W_{k,k}\\
W_i&=& \frac 1{\vec \nabla ^2} (h_{ik,k} -\frac 14 h_{kk,i}-\frac 14
\frac 1{\vec \nabla ^2} h_{kl,kli})
\end{eqnarray}
\end{subequations}
It can be checked that at linear order $h^{TT}_{ij}$ defined by York
equals that of ADM. Moreover, all gauge dependence is contained in
the pure-gauge part $W_{i,j} +W_{j,i}$ in $\tilde h^L_{ij}$, while
the $ -\frac 23 \delta _{ij} W_{k,k}$ term in $\tilde h^L_{ij}$ can
join $\frac 13 \delta _{ij} h_{kk}$ to make a gauge-invariant
combination:
\begin{equation}
\frac 13 (h_{kk}-2 W_{k,k})=\frac 12 (h_{kk}-\frac 1{\vec \nabla^2}
h_{kl,kl})=\frac 12 f^T.
\end{equation}

It must be noted, however, that the $W_i$ of York differs from the
$f_i$ of ADM, and the pure-gauge terms defined by York and ADM are
different: $W_{i,j} +W_{j,i}\neq f_{i,j}+f_{j,i}$. They are both
much more complicated than our pure-gauge term in Eq. (\ref{hatgs}):
\begin{subequations}
\begin{eqnarray}
\bar h^{(1)}_{\mu\nu}&=&\frac 1{\vec\nabla ^2}(h_{\mu i,i\nu}+h_{\nu
i,i\mu} -h_{ii,\mu\nu}) \\
&=&\epsilon_{\mu,\nu} +\epsilon_{\nu,\mu}, ~ \epsilon_\mu = \frac
1{\vec\nabla ^2}(h_{\mu i,i}-\frac 12 h_{ii,\mu}).
\end{eqnarray}
\end{subequations}

The relations between our decomposition and that of ADM and York,
especially beyond the linear order, will be further explored
elsewhere. \cite{Chen11}

This work is supported by the China NSF Grants 10875082 and
11035003. XSC is also supported by the NCET Program of the China
Education Department.


\begin{thebibliography}{99}
\bibitem{note1} Greek indices run from 0 to 3, Latin
indices run from 1 to 3, and repeated indices are summed over (even
when they both appear upstairs or downstairs).
\bibitem{noteBound} For an infinite system, it is not obvious to
specify a natural boundary condition. Though an infinite system is
hardly relevant for particle physics, in gravity the universe does
present a real example. Here we restrict out attention to finite
systems. Note that this does not exclude the radiating system, which
can be made spatially finite by letting the radiation occur in a
given period.
\end{thebibliography}
\end{document}